\RequirePackage{etoolbox}
\csdef{input@path}{{style/}{graphics/}}
\documentclass[aoas,MSNbibl,nameyear,dvips]{arximspdf}

\doi{10.1214/14-AOAS790} 
\volume{8}
\issue{4}
\pubyear{2014}
\firstpage{2567}
\lastpage{2567}
\docsubty{FLA}


\begin{document}
\begin{frontmatter}

\title{Acknowledgment of priority
Usage~of~the~Lambert~W~function~in~statistics}
\runtitle{Acknowledgment of priority}
\pdftitle{Acknowledgment of priority: Usage of the Lambert $W$ function in statistics}

\begin{aug}
\author[A]{\fnms{Georg M.}~\snm{Goerg}\corref{}\ead[label=e1]{im@gmge.org}}
\runauthor{G. M. Goerg}
\affiliation{Google, Inc.}
\address[A]{Google, Inc.\\
111 8th Avenue\\
New York, New York 10011\\
USA\\
\printead{e1}}
\end{aug}

\received{\smonth{9} \syear{2014}}
\revised{\smonth{10} \syear{2014}}


\end{frontmatter}

In my 2011 \textit{Annals of Applied Statistics} article [\citet
{GMGLambertWSkewed}] I wrote that ``Whereas the Lambert $W$ function
plays an important role in mathematics, physics, chemistry, biology and
other fields, it has not yet been used in statistics.'' This was
incorrect. At the time of publication I was unaware of \citet
{Stehlik03LambertWLoglikTest}, who used the Lambert $W$ function to
derive the exact distribution of the likelihood ratio test statistic.
He has also used it in more recent work such as \citet
{Stehlik06exactLR,Stehliketal10favorable}, or
\citet{Stehlik14exacttesting} amongst others. While Stehl\'{i}k's
use of the
Lambert $W$ function was focused on the distribution of the likelihood ratio
test statistic, my work dealt with the modeling of skewed random
variables and symmetrizing data
using the Lambert $W$ function as a variable transformation.

I thus want to take this opportunity to acknowledge that Stehl\'{i}k
has used the Lambert $W$ function prior to my usage in \citet
{GMGLambertWSkewed}.

%

\printaddresses
\end{document}